\documentclass[aps,preprint,pra,showpacs]{revtex4}%
\usepackage{graphicx}
\usepackage[pdftex]{color}
\usepackage{amssymb}
\usepackage{amsfonts}
\usepackage{amsmath}%
\providecommand{\U}[1]{\protect\rule{.1in}{.1in}}
\def\be{\begin{equation}}
\def\ee{\end{equation}}
\def\bea{\begin{eqnarray}}
\def\eea{\end{eqnarray}}

\begin{document}
\title{Is the Quilted Multiverse Consistent with a Thermodynamic Arrow of Time?}
\author{Yakir Aharonov}
\affiliation{School of Physics and Astronomy, Tel Aviv University, Tel Aviv 6997801, Israel}
\affiliation{Schmid College of Science, Chapman University, Orange, CA 92866, USA}
\author{Eliahu Cohen}
\affiliation{H.H. Wills Physics Laboratory, University of Bristol, Tyndall Avenue, Bristol
BS8 1TL, UK}
\author{Tomer Shushi}
\affiliation{Department of Physics, Ben-Gurion University of the Negev, Beersheba 8410501, Israel}
\date{\today}

\begin{abstract}
Theoretical achievements, as well as much controversy, surround multiverse
theory. Various types of multiverses, with an increasing amount of complexity,
were suggested and thoroughly discussed in literature by now. While these types are very
different, they all share the same basic idea: our physical
reality consists of more than just one universe. Each universe within a
possibly huge multiverse might be slightly or even very different from the
others. The quilted multiverse is one of these types, whose uniqueness arises
from the postulate that every possible event will occur infinitely many times
in infinitely many universes. In this paper we show that the quilted
multiverse is not self-consistent due to the instability of entropy decrease
under small perturbations. We therefore propose a modified version of the
quilted multiverse which might overcome this shortcoming. It includes only
those universes where the minimal entropy occurs at the same instant of
(cosmological) time. Only these universes whose initial conditions are
fine-tuned within a small phase-space region would evolve consistently to form
their ``close'' states at present. A final boundary condition on the
multiverse may further lower the amount of possible, consistent universes.
Finally, some related observations regarding the many-worlds interpretation of
quantum mechanics and the emergence of classicality are discussed.

%\textit{Keywords: Multiverse theory, quilted multiverse, arrow of time,
%stability, many-worlds interpretation }

\end{abstract}

\pacs{05.70.-a, 98.80.-k}
\maketitle

\section{Introduction}

Multiverse theory (also known as Meta universe theory) is a group of models
assuming that our physical reality encompasses more than one universe, i.e.
there exists at least one more universe other than ours. Several types of such
multiverses are known in literature \cite{A1,C1,C3,J1,R1,W1,W2,G4,G1,G3,G2,T2}.

Some of these models suggest that our physical reality comprises of infinitely
many universes \footnote{In the context of this work we shall assume a discrete phase-space, meaning that all infinities are countable.}, while others postulate that we live in a multiverse with a
finite number of universes. Most multiverse theories imply that universes
might not be uniquely identified through their macroscopic state at present or
past, i.e. their macrostates could be quite similar during long and even
infinite time intervals. The ultimate multiverse model (also known as the
mathematical multiverse) \cite{T1} satisfies this property and postulates that
every possible state is in one-to-one correspondence with each universe from
the multiverse horizon.

One of the most common explanations of the big-bang is given by quantum
fluctuation theory, which suggests that our universe began from a quantum
fluctuation, and if so, it is natural to deduce that in our physical reality
these fluctuations are taking place in all of our space and time dimensions
(see \cite{G4} for instance). Therefore, an infinite number of such
fluctuations implies a vast multiverse of infinite number of universes.

The multiverse type that we shall focus on is the quilted multiverse
\cite{G2}, whose infinite space and time dimensions presumably contain
infinite number of universes. In Greene's words \cite{G2}: ``At any moment in time, the
expanse of space contains an infinite number of separate realms-constituents
of what I'll call the Quilted Multiverse-with our observable universe, all we
see in the vast night sky, being but one member. Canvassing this infinite
collection of separate realms, we find that particle arrangements necessarily
repeat infinitely many times. The reality that holds in any given universe,
including ours, is thus replicated in an infinite number of other universes
across the Quilted Multiverse.''

The quilted multiverse provides a theoretical probabilistic approach for the
existence of events before the event horizon in our physical reality. Within
the quilted multiverse, the event horizon includes events that occur
infinitely many times, duplicated in infinitely many universes, which might be
finite or infinite. From the characterization above we deduce that there are universes within the
quilted multiverse that are not only ``close'' at a given time (e.g. at
present), that is, similar in a sense that will be defined below, but have
been very ``close'' for a substantially large time interval.
In terms of Tegmark's hierarchy \cite{T2}, the quilted
multiverse we shall study corresponds to a level 1 multiverse. This type of
multiverse postulates that every universe in the multiverse shares the same
physical constants (e.g. the Planck constant $\hbar$ and the speed of light
$c$), while other types of multiverses suggest that the physical constants and
even physical laws are different within different universes (e.g. string
theory landscape \cite{Sus}). The main argument for this kind of multiverse
with different physical constants is that for different universe we would have
different spontaneous symmetry breaking and thus different physics. Since
there are already several arguments against this type of multiverse (see
\cite{W4} for instance), we will focus in this paper on the quilted multiverse
where parallel universes share the same physical constants and same physical
laws. We also emphasize that the quilted multiverse differs from the
inflationary multiverse. The former emerges if the extent of space is
infinite, while the latter's variety emerges from eternal inflationary expansion.
We would assume that the multiple universes within the quilted multiverse can be coarse-grained in a countable manner, they have the same common cosmological features and same local laws, and they do not interact with each other.

We claim in this work that the quilted multiverse is not consistent with basic
thermodynamic assumptions. In the following section we discuss a thermodynamic
arrow of time defined by the stability of entropy increase. In
section \ref{sec4} we present an inconsistency of the quilted multiverse and
the proposed thermodynamic arrow of time. Section \ref{sec5} discusses an
upper bound to the number of parallel universes in the quilted multiverse, and
section \ref{sec6} attempts to broaden these results towards other kinds of multiverse when adding to the analysis a final boundary condition. Section \ref{sec7} concludes the paper.

\section{A subtle thermodynamic arrow of time}

\label{sec2}

\label{arrow_of_time} Time seems to incessantly ``flow'' in one direction,
raising the ancient question: Why? This intensively discussed question can be
answered in several ways by introducing seemingly different time arrows:
thermodynamic, cosmological, gravitational, radiative, particle physics
(weak), quantum and others \cite{Savitt,Zeh}. We employ in this paper the
cosmological arrow of time, which points in the direction of the universe's
expansion. This choice implies that parallel universes with the same
macrostate will have the same time. Our main argument, however, will rely on
thermodynamic stability under small perturbations which allows to define
another crucial arrow of time -
\textit{the macroscopic behavior of a large system is stable against
perturbations as far as its future is concerned, but for most cases is very unstable as far as
its past is concerned.} \cite{L1,L2}. The positive direction of time is thus
determined according the system's stability under small perturbations. Indeed,
performing a slight microscopic change (not to mention a large macroscopic change) in the system's past will not change, in general, its macrostate at future times, i.e. the system will end up its
time evolution with the same high entropy macrostate. However, when
propagating backwards in time, such a slight change in the system's future
will have far-reaching consequences on its past \cite{L1,L2}. This is the key
observation we shall utilize next, akin to the thermodynamic arrow of time which relays on the second law of thermodynamic (although some subtle challenges are known in literature \cite{2ndLaw1,2ndLaw2}). The difference in terms of stability
between future and past stems from the fact that any perturbation of a
microstate will tend to make it more typical of its macrostate and thus small
perturbations will not interfere with (forward in time) typical evolution.
Backwards in time, however, the microstate will propagate towards a smaller
phase space volume which is untypical of the macrostate. This difference in
Lyapunov stability was rigorously quantified, e.g. in \cite{Hoover,Sarmen}.

In this perspective paper we will formally treat a universe parallel to ours, having at
present time a similar macrostate or even the same macrostate, yet with a
slightly different microstate as a perturbation. Then we will try to apply the
above thermodynamic reasoning.

\section{Inconsistency of the quilted multiverse}

\label{sec4}

Before we claim that the quilted multiverse is inconsistent with the
instability of entropy decrease discussed in Sec. \ref{arrow_of_time}, let us
define some mathematical symbols which will be useful later on. First, suppose
that we have an infinite (yet countable) number of universes, denoted by
\begin{equation}
\label{U111}\mathcal{U=}\{\mathcal{U}_{1}, \mathcal{U}_{2},...\},
\end{equation}
where each universe $\mathcal{U}_{j}$ has the (quantum) microstate $\Psi_{j}(t) $,
and $t$ is the cosmological time.

Further, let us define in phase space a distance measure $\Delta$, which quantifies the
difference between the microstate of the $j-$th universe, $\Psi_{j}(t),$ and
the microstate of the $i-$th universe, $\Psi_{i}(t),$ at some time $t_{i}
=t_{j}=t$
\begin{equation}
\Delta(\Psi_{j}(t),\Psi_{i}(t))>0,
\end{equation}
for $i\neq j$. We hereby define $0 \le \Delta \le 1$ to be the ratio between the number of particles whose (possibly entangled) states are orthogonal and the total number of particles. This definition might not be unique (or the most robust) but it captures our intuition as to microscopic proximity of similar/identical macroscopic states.
According to this definition and Greene's description of the
quilted multiverse, for every $\varepsilon>0$ there exist at least two
universes such that
\begin{equation}
\Delta(\Psi_{j}(t),\Psi_{i}(t))\leq\varepsilon,\text{ }\forall t\in T,
\end{equation}
where $T=[t_{0},t_{f}]$ is some long time interval comparable with the age of
the universes.

Moreover, from the above description of the quilted multiverse, we deduce that every
possible event will occur an infinite (countable) number of times. Therefore, this model
suggests that there should exist a set $W$  such that
\begin{equation}
\label{W111}W=\{(i,j)|\Delta(\Psi_{j}(t),\Psi_{i}(t))\leq\varepsilon,\forall t\in
T\}, |W|=\aleph_0
\end{equation}
where $\varepsilon$ is some threshold below which we may say that the universes are ``close'', and $W$ is the set of all
possible pairs of universes from the infinite multiverse (\ref{U111}) that are ``close'' for a period of time comparable with their age.

\begin{figure}[h]
\centering \includegraphics[height=10cm]{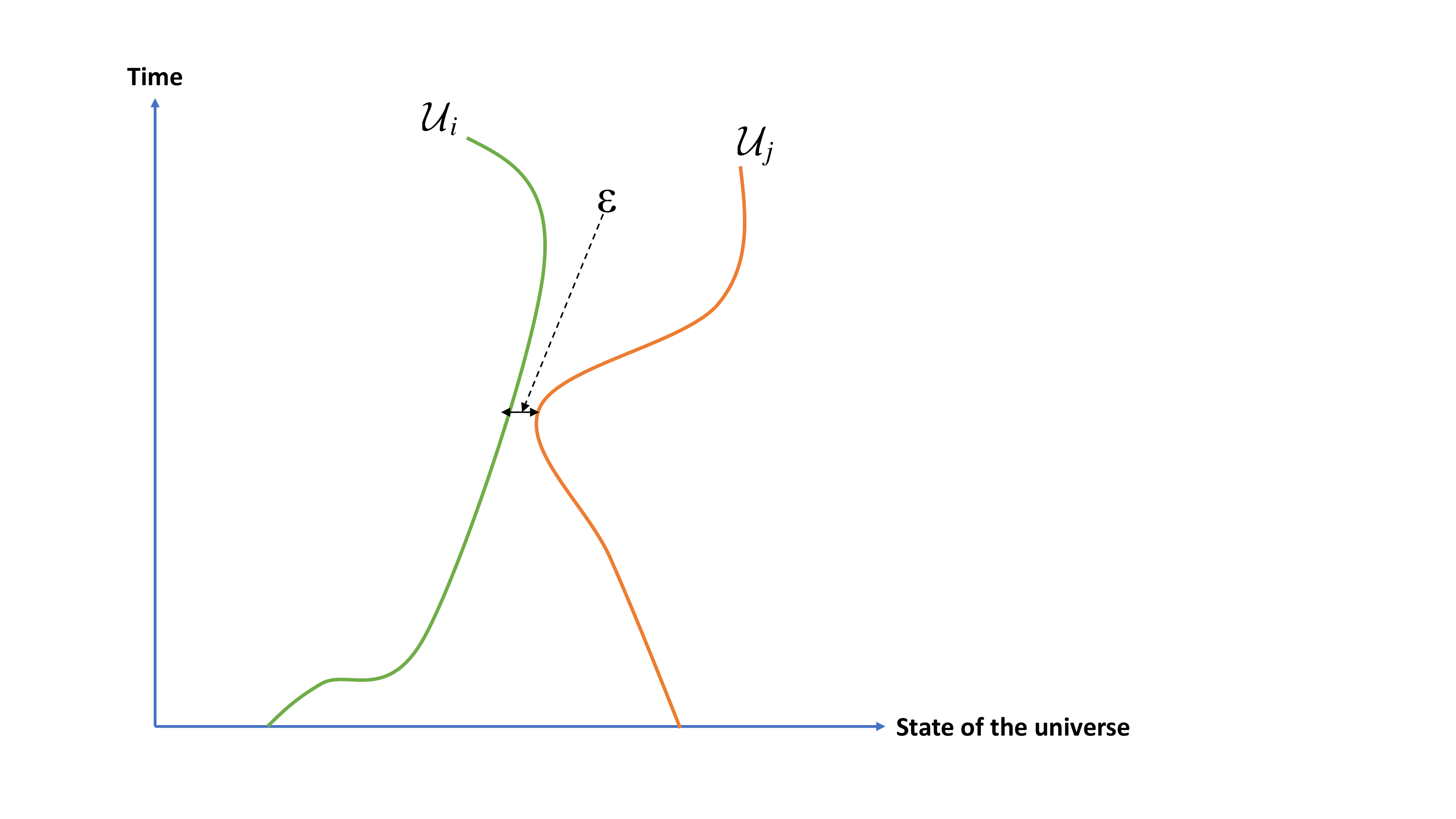}\caption{Two ``close''
universes at present time were most likely ``far'' in the past.}
\label{fig1}
\end{figure}

We now show that (\ref{W111}) is not consistent with the thermodynamic arrow of
time defined in Sec. \ref{arrow_of_time} (it will be implicitly assumed that the universe is in a non-equilibrium state). First, notice
that if we have a thermodynamic system $\mathcal{V}$ with the macroscopic state
$M_{\mathcal{V}}(t)$ at the cosmological time $t$, there exist more than
one possible quantum state $\Psi_{\mathcal{V}}(t_0)$ (for every $0<t_0<t$) that will
produce $M_{\mathcal{V}}(t)$ in time $t,$ i.e. there is some volume in the past phase space of (quantum) microscopic states that could reproduce the present macroscopic state. However, a slightly different universe at present (analogous to a small perturbation of the first) would correspond in general to a very different volume in the past phase space \cite{L1,L2}, which would in turn correspond to a markedly different macroscopic state for all times. Therefore, the backwards evolution in time (presumably dictated by the same dynamical rules) of two very close universes at present will result in two very far universes at the past (see Fig. \ref{fig1}), thus negating
(\ref{W111}).

There is only a negligible probability that two close
universes at present, will evolve backwards in time to two close universes in
the past (see also \cite{ABE}).

%This can be stated mathematically as follows:

%If
%\begin{equation}
%\Delta(\Psi_{i}(t),\Psi_{j}(t))\leq\varepsilon,
%\end{equation}
%then, it is not true in general that $\Delta(\Psi_{i}(t-\alpha),\Psi
%_{i}(j-\alpha))\leq\varepsilon,$ for an arbitrary $\alpha>0.$ Hence, the claim
%that many universes approximately share both our present and past (see
%\cite{G2} for instance), is not consistent with our notion of a thermodynamic
%arrow of time.

Also, it is inconsistent to assume that any arbitrary change to our current universe is a valid parallel universe having the same historical source in phase space or having the same point in time of minimum entropy.

%This is an outcome of the following fundamental fact discussed in Sec.
%\ref{arrow_of_time}: the time direction in which entropy grows is stable
%against minor changes at a given $t$, while the opposite direction is highly unstable.

%Furthermore, the entropy corresponding to our universe at present time is most
%likely to be the minimal one; it will grow in both directions of time,
%allowing $\Delta$ to be small only for very short time intervals.

\bigskip The number of possible universes can be represented by the Boltzmann
relation between entropy $S$ and the set $\Omega$ of possible microstates
corresponding to the same macrostate,
\begin{equation}
S=k_{B}\ln|\Omega|,
\end{equation}
where $k_{B}$ is the Boltzmann constant.

\bigskip Then, given the entropy of the i-th universe, $S_{\mathcal{U}_{i}},$
$|\Omega_{\mathcal{U}_{i}}|$ is
\begin{equation}
|\Omega_{\mathcal{U}_{i}}|=e^{S_{\mathcal{U}_{i}}/k_{B}}.
\end{equation}
Assuming that during its 13.8 billion years of history the universe has
reached a very large entropy $S_{\mathcal{U}_{i}} \gg k_{B}$, we have a huge set
of possible microstates $\Omega_{\mathcal{U}_{i}}.$ Let us examine a pair of
universes having at $t$' close states, i.e. $\Delta(\Psi_{i} (t^{\prime}
),\Psi_{j}(t^{\prime}))\simeq0$, where $\Psi_{i}\in\Omega_{\mathcal{U}_{i}}$,
$\Psi_{j}\in\Omega_{\mathcal{U}_{j}}$ and $|\Omega_{\mathcal{U}_{i}}
|=|\Omega_{\mathcal{U}_{j}}|$. We claim that at arbitrary time
$t^{\prime\prime}\ll t^{\prime}$ they will most likely have $|\Omega
_{\mathcal{U}_{i}}|\neq|\Omega_{\mathcal{U}_{j}}|$, and the probability that
$\Delta(\Psi_{i}(t^{\prime\prime}),\Psi_{j}(t^{\prime\prime}))\simeq0$ will be
close to zero. This follows from the fact that $\Omega_{\mathcal{U}_{k}
}(t^{\prime\prime})$ is now the backward-in-time evolution of $\Omega
_{\mathcal{U}k}(t^{\prime})$, $k=i,j,$ which is a set with very large number
of possibilities, so that the probability
\begin{equation}
\Pr\left(  |\Omega_{\mathcal{U}_{i}}(t^{\prime\prime})|\simeq|\Omega
_{\mathcal{U}_{j}}(t^{\prime\prime})|\mid|\Omega_{\mathcal{U}_{i}}(t^{\prime
})|\simeq|\Omega_{\mathcal{U}_{j}}(t^{\prime})|\right)  ,
\end{equation}
will be zero.

Another way to see this inconsistency is to consider the point of minimal
entropy during the lifetime of our universe. When picking at random another
hypothetical universe having at present the same macrostate as our universe,
it is most likely to have its minimal entropy at some other time different
from ours (most likely after ours). Hence, the histories of the two universes
cannot be the same, unless we focus at present only on the zero measure of
macrostates having their minimal entropy at exactly the same time as ours.

\section{Upper bound to the number of parallel universes in the quilted
multiverse}

\label{sec5}

We shall try to approach the problem from a different perspective now,
beginning with some qualitative considerations. One should note two extreme
distance scales between universes in a multiverse. When two universes are extremely close (that is, different but virtually indistinguishable so that $0 < \Delta \ll 1$) at some point in
time, they may have a non-negligible probability evolving backwards to extremely close initial states, thereby
creating no inconsistency. However, having infinitely many universes which are identical to ours for all practical purposes is not too interesting. On the other hand, if two universes are far apart right now, stability (which corresponds to small perturbation) again plays no role. But this is not the case we wish to rule out.

Between these two contingent cases, lie the problematic distances to
which instability considerations can be applied. This may pose a constraint on
the distribution of universes within a multiverse - there might be infinitely many
universes which are very far from each other and an infinite number of universes which are
extremely close, but we do not expect too many universes to be intermediately
close when we demand consistency over long times.

Let us examine now for concreteness a $6N$-dimensional quantum phase space. Let us suppose for simplicity that the phase space is discrete and
focus on some large yet finite part of it. We can therefore think about this
sector as a hypercube with a fine grid. A universe $j$ within this sector is
represented e.g. by a Wigner quasi-distribution $W\left(  X_{j},P_{j}\right)  $ on the
grid, where $X_{j}$/$P_{j}$ encapsulates the three position/momentum vectors,
respectively, of each particle in this universe. We now start to gradually
fill the hypercube with more and more distributions. We begin with those having a slight overlap (or no overlap at all) with the original one and with each other, thus corresponding to universes which are very different. As this process continues, we will have to fill the finite phase space with more and more distributions, closer to each other, until a point (let us denote it by $\Delta=D$) when they become very close, such that distance between the universes characterizes a small perturbation. We would thus unavoidably create at least two universes that are too close to each other. Too close, in the sense that one can be thought of as a small perturbation to the other, and
then upon backward evolution in time, they would most likely reach inconsistent states.

We now apply similar arguments to those appearing at the end of the previous
Section. It seems that in a countably infinite phase space (allowing a countably infinite number of parallel universes) and a finite point in time $t$, there might be only a finite number of consistent parallel universes whose $\Delta$ separation is very close until time $t=0$, but we leave this as a conjecture. In any case, we would like to point out that an infinite number of parallel universes might be ruled out this way just as a result of thermodynamic considerations.

\bigskip

To resolve this apparent shortcoming of the quilted multiverse we must pose a condition on the possible
distance between the universes, and eventually on their density. In case
that
\begin{equation}
\Delta\left(\Psi_{i},\Psi_{j}\right)  \geq D,~\forall~i,j
\end{equation}
for some threshold $0<D<1$, we potentially find a consistent multiverse that does not
violate the aforementioned notion of stability. To this microscopic condition we add the macroscopic demand that despite the distance, the various universe would still describe the same macrostate at all times, and in particular would have their minimal entropy state at the same cosmological time. Of course, this multiverse is different
from the quilted multiverse, and hence we call it the \textquotedblleft
Modified quilted multiverse\textquotedblright. As opposed to the ordinary
quilted multiverse, it might coexist with thermodynamic laws, yet may still violate other basic requirements like Occam's razor \footnote{In the quilted multltiverse, the number (or commonness) of universes does not correspond to probability/Born rule, but in contrast, for the many-worlds-type multiverse we would have to employ a different logic as presented in Sec. \ref{sec6}.}.

\section{Generalizations employing a final boundary condition}
\label{sec6}

It could be interesting to apply the above considerations to other kinds of
multiverses. However, when the values of physical constants, and moreover,
physical laws themselves, in other universes become different from those we
know now in our universe, the distance between our universe and others might
be very large at present (and furthermore vary with time). Therefore, it is
not obvious how to apply stability considerations to these kinds of multiverse.

On the other side of the multiverse scale, there is the many worlds
interpretation of quantum mechanics (also known as the quantum multiverse). In
previous works \cite{ACL1,AC,ACL2}, two of us have employed a final boundary
condition on the universe which is of special kind. This unique boundary
condition allowed us to overcome some conceptual difficulties appearing in the
many worlds interpretation. In particular, we suggested a model for a macroscopically reversible universe without the need of employing infinitely many parallel universes. Furthermore, we were able to devise an effective
collapse mechanism in this single-branched ``modest'' multiverse structure. Finally, our proposed two-time decoherence scheme allowed to draw the boundary between the classical and quantum regimes.

These past results hint that the multitude of universes proposed by the many-worlds interpretation may not be needed in order to account for our empirical observations in a time-symmetric manner. Other kinds of multiverse can be handled the same way, and indeed, posing both initial and final boundary conditions on a multiverse should dramatically lower the measure of possible universes within it: Regardless of the dynamics,
when the final state of the multiverse is evolved backwards in time, it must
be compatible with any earlier state. As noted in \cite{AR}, some final
boundary conditions give rise to the Born rule, and are hence preferable over
others. Further conditions on the final state may even isolate a unique set of
final boundary conditions with a higher explanatory power. These include our
proposal for a quantum universe having a natural notion of classicality
emerging from the requirement to store microscopic information in a redundant
manner \cite{ACL1,AC,ACL2}. A recently analyzed feature of this time-symmetric universe is a top-down logical structure \cite{ACT}, which could further shed light on the subtle relations between micro and macro scales.

\section{Conclusions}

\label{sec7}

Multiverse theory has various models that describe different structures of the
physical reality. One of these models is the quilted multiverse, which
postulates that every possible event is occurring infinitely many times in
nature, thus there are infinitely many universes resembling ours. At first
glance, this model seems to be self-consistent. However, we have shown that
this model negates basic thermodynamic principles. The difference between
microstates in two ``close'' universes cannot be $\epsilon$ small at each
point in time, or even along a finite, sufficiently large time interval.
Therefore, any possible type of multiverse would better not assume such a
relation between two universes. Moreover, every universe must have its unique
past and future in the sense that there is no other universe with the same, or
even very close, state over a substantial part of its life time. We therefore
have to limit ourselves only to those universes whose macrostates at present
time evolve backwards to the same point in time of minimal entropy such as
ours. These obviously reside in a very small fraction of phase space
and may evolve in a consistent way. Further constraints on the number of
possible universes may arise when augmenting this analysis with a final
boundary condition on the multiverse.

These findings corroborate previous ones of our group \cite{ACL1,AC,ACL2}, suggesting that in
addition to apparent inconsistencies and various conceptual problems, the overwhelming multitude exhibited by
multiverse theory in general, and the quilted mutliverse/many worlds
interpretation in particular, might not be needed in order to satisfactorily account for our
observations in the classical and quantum realms using a single, unique universe.

\section*{Acknowledgements}

We wish to thank Avshalom C. Elitzur and Daniel Rohrlich for helpful comments.
Y.A. acknowledges support of the Israel Science Foundation Grant No. 1311/14,
of the ICORE Excellence Center ``Circle of Light'' and of DIP, the
German-Israeli Project cooperation. E.C. was supported by ERC AdG NLST. T.S.
thanks the John Templeton Foundation (Project ID 43297) and from the Israel
Science Foundation (grant no. 1190/13). The opinions expressed in this
publication are those of the authors and do not necessarily reflect the views
of any of these supporting foundations.

\end{document}